\documentclass[12pt,aps,pra,preprint,floatfix]{revtex4} 
\usepackage{amssymb,epsf}

\usepackage{latexsym}
\usepackage{xcolor}
\usepackage{epsfig}
\usepackage{float} 
\usepackage{color} 
\usepackage{amsmath}
\usepackage{bm}
\usepackage{amsthm}
\usepackage{amssymb}
\usepackage{amssymb,epsf}
\usepackage{latexsym}
\usepackage{epsfig}
\usepackage{graphicx}

\begin{document} 

\title{Quantum gravity inspired nonlocal quantum dynamics preserving the classical limit}
\author{Marzena Ciszak}
\affiliation{CNR-Istituto Nazionale di Ottica c/o LENS, I-50019 Sesto Fiorentino (FI), Italy.}
\author{Alessio Belenchia}
\affiliation{Institute of Quantum Technologies, German Aerospace Center (DLR), D-89077 Ulm, Germany}
\author{Antonello Ortolan}
\affiliation{INFN, Laboratori Nazionali di Legnaro, I-35020 Legnaro (Padova), Italy}
	\author{Francesco Marino \footnote{Corresponding author}}
	\email{francesco.marino@ino.cnr.it}
	\affiliation{CNR-Istituto Nazionale di Ottica c/o LENS and INFN, Sezione di Firenze, I-50019 Sesto Fiorentino (FI), Italy.}

\begin{abstract}
Several approaches to quantum gravity lead to nonlocal modifications of fields' dynamics. This, in turn, can give rise to nonlocal modifications of quantum mechanics at non-relativistic energies. Here, we analyze the nonlocal Schr\"{o}dinger evolution of a quantum harmonic oscillator in one such scenario, where the problem can be addressed without the use of perturbation theory. 
We demonstrate that although deviations from standard quantum predictions occur at low occupation numbers, where they could potentially be detected or constrained by high-precision experiments, the classical limits of quantum probability densities and free energy remain unaffected up to energies comparable with the nonlocality scale. These results provide an example of nonlocal quantum dynamics compatible with classical predictions, suggesting massive quantum objects as a promising avenue for testing some phenomenological aspects of quantum gravity.


\end{abstract}
\maketitle

\section{Introduction}

Theories preserving Lorentz invariance while accounting for the emergence of the continuum spacetime from more fundamental constituents typically involve nonlocal equations of motion. Relevant examples in the quantum gravity (QG) literature include causal set theory, where the interplay between Lorentz invariance and discreteness results in nonlocal dynamics for fields residing on the causal set \cite{sorkin,belen15}, string theory and string field theory, where the string and its interactions inherently exhibit nonlocality \cite{eliezer,taylor}, and noncommutative geometry \cite{szabo}. Similar forms of nonlocality are expected to arise in loop quantum gravity in order to avoid significant violations of Lorentz invariance, pointing to an interesting connection with string theory physics \cite{gambini}. 

Interestingly, imposing Lorentz invariance and requiring the avoidance of instability according to the Ostrogradsky theorem~\cite{ostro} restricts the possible modifications of the dynamics singling out the ones with infinitely many derivatives, i.e., nonlocal~\cite{barnaby2008dynamics}. Furthermore, the scale that characterizes the nonlocality does not necessarily correspond to the Planck scale, as it is indeed the case in the QG examples mentioned. These observations have stimulated experimental proposals aimed at constraining this mesoscopic scale~\cite{biswas,belen16,PhysRevD.94.061902}.

Motivated by ongoing experiments involving quantum optomechanical oscillators, a non-local Klein–Gordon field in the non-relativistic limit was derived and the corresponding non-local Schr\"{o}dinger equation was solved perturbatively ~\cite{belen16}. This analysis, further developed in ~\cite{belen17,belen19}, suggests that table-top experiments with massive quantum objects have the potential to improve the experimental constraints of non-local field theories currently based on LHC’s data~\cite{biswas}. 
However, the results of Refs.~\cite{belen16,belen17,belen19} were based on a perturbative expansion of the nonlocal Schr\"{o}dinger equation, thus making the dynamics intrinsically local. To what extent the first-order approximation is able to reproduce the key features of the non-local dynamics is yet to be fully understood. For instance, when applied to the paradigmatic case of a harmonic oscillator, nonlocality, at the lowest order in the perturbation parameter, introduces additional Hamiltonian terms~\cite{belen19} which result in an energy spectrum that scales quadratically with the occupation number. 
A related important aspect concerns the form of the classical limit corresponding to the modified quantum dynamics.
On the one hand, there is a wide consensus that these modifications should manifest themselves just on purely quantum systems. We observe indeed that decoherence, which plays a major role in determining the emergence of classical behaviour as a limit of a quantum description is not usually included in phenomenological quantum gravity models. Hence, it wouldn't be surprising if some features predicted by considering the intertwined influences of gravity and quantum physics were washed out, for instance, for systems in a thermal state. On the other hand, if the modified dynamics apply consistently to states of high purity, their validity should extend equally to both the quantum ground state and high-order Fock states. Incompatibility with the classical predictions at occupancy $n \rightarrow \infty$ would raise a number of conceptual issues, limiting \textit{de facto} the reliability of these models. 
The problem of reproducing sensible classical limits starting from deformed quantum rules is common to
many phenomenological approaches to quantum gravity. For instance, in the framework of spacetime noncommutativity the nonlinear law of addition of momenta, necessary to preserve Lorentz invariance, might inevitably produce a pathological description of the total momentum of a macroscopic body~\cite{hossenfelder}.

In this work, we study the evolution of a quantum harmonic oscillator in the framework of a string-inspired nonlocal model. While the dynamics is significantly affected at low occupancy, we show analytically and verify numerically that the classical limit of the quantum probability density is recovered. From the numerically calculated energy eigenvalues we also show that the Helmholtz free energy tends to the standard classical value at high occupation numbers. These results provide an example of a nonlocal quantum dynamics preserving the classical limit. While our analysis mainly focuses on a specific type of nonlocality, we argue for the generalization of our results to other cases of physical relevance and, more importantly, the applicability of our approach as a consistency check. Any model whose nonlocal function fails to restore the classical limit should be promptly ruled out. 

The paper is organized as follows. In Sec. II we provide a brief overview of the theoretical framework that forms the basis for the model under examination. For additional details, we direct readers to Refs. \cite{belen16,belen17,belen19}. In Sec. III, we provide a consistent definition of classical limit in the presence of a nonlocality scale and we show that at high energies the nonlocal quantum probabilities gradually tend towards this limit. In Sec. IV we calculate the quantum Helmholtz free energy and show that even for this quantity the classical limit is preserved. In Sec. V, we discuss the generality of the argument and argue for its validity beyond the specific functional form considered. Our conclusions are given in Sec. VI.  

\section{Nonlocal model of a quantum harmonic oscillator}

The general idea is that the evolution equation of any relativistic field ($\phi$), encoding a covariantly defined scale $l_k$ characteristic of the elementary spacetime constituents and preserving Lorentz invariance, must necessarily take the form $f(\Box+\mu^2)\phi=0$, where $\Box = -c^{-2}\partial_t^2+\nabla^2$ is the d'Alembert operator and $1/\mu=\hbar/m c$ is the reduced Compton wavelength of the field \cite{belen16,belen17}. Here, $f$ is some non-polynomial function such that $f(\Box+\mu^2)\rightarrow \Box+\mu^2$ in the limit $l_k\rightarrow 0$. The generalized Klein-Gordon operator so defined is inherently nonlocal as it contains an infinite number of both temporal and spatial derivatives. This feature is essential to avoid the instabilities prescribed by the Ostrogradski's theorem \cite{ostro}. It is natural to expect the scale $l_k$ to enter the low energy physics as a perturbative parameter of the local evolution. The function $f$ can therefore be interpreted as providing the UV completion of the standard local theory and $l_k$, which we refer to as the nonlocality length scale, as the scale at which corrections to the standard evolution come into play. It is important to remark that this scale does not necessarily have to be related to the spatiotemporal discreteness normally associated to the Planck scale, a fact which is of particular relevance within the context of casting phenomenological constraints. In this framework, $l_k$ is considered as a free parameter of the theory, to be bound by the experiments.

In this section, we outline how such non-local effective field theories lead to a modified Schr\"{o}dinger evolution in the nonrelativistic limit and then focus on the specific case of a quantum harmonic oscillator in the framework of a string-inspired nonlocal dynamics.

\subsection{Non-relativistic limit of nonlocal effective field theories}

We consider a free complex massive scalar field, $\phi$, of mass $m$ defined by the Lagrangian $\mathcal{L} = \phi(x)^*f(\Box+\mu^2)\phi(x)+c.c.$. Following standard treatments (see e.g. \cite{tong2006lectures}) we decompose the field as $\phi(x)=e^{-i \frac{m c^2}{\hbar}t}\psi(t,x)$. Substituting this into the Lagrangian $\mathcal{L}$ and taking the non relativistic limit ($c\rightarrow\infty$), we find
\begin{equation}
\mathcal{L}_{\mathrm{NR}}  =  \psi^{*}(t,x)f(\mathcal{S})\psi(t,x)+c.c.,
\label{lag2}
\end{equation}
where $\mathcal{S}= i\hbar \partial_t + \frac{\hbar^2}{2m}\nabla^2$ is the usual Schr\"odinger operator. One can also include an external potential, $V(x)$, by adding the term $V(x)\psi^*\psi$ in Eq.~
(\ref{lag2}).

The equations of motion can be derived by means of a nonlocal generalisation of the Euler-Lagrange equations~\cite{bollini} which gives 
\begin{equation}
f(\mathcal{S})\psi(t,x)= V(x) \psi(t,x) \, .
\label{nls}
\end{equation}

The nonrelativistic complex field $\psi$ satisfies a nonlocal generalisation of the Schr\"odinger equation. In order to interpret $\psi$ as the wavefunction of a quantum mechanical system one should construct from the field operator $\psi$ a wavefunction for a generic one particle state and show that it satisfies the same Schr\"odinger equation as the field. While this is indeed the case at the first-order of a perturbative expansion of $f(\mathcal{S})$~\cite{belen19}, the generalization of this result to the full theory would require the derivation of the nonlocal Hamiltonian, a result that is technically challenging. Thus, from here on we will proceed with the caveat that our approach is phenomenological, in the sense that a rigorous identification of $\psi$ with the wavefunction of a quantum mechanical system is yet to be established.

\subsection{Non-local Schr\"{o}dinger evolution in a harmonic potential}

The functional form of $f(\mathcal{S})$ depends on the specific quantum gravity model under consideration. Here we focus on a nonlocal Klein-Gordon operator of the form
\begin{equation}
f(\Box+\mu^2)=(\Box+\mu^2)\,\exp{[l_k^2(\Box+\mu^2)]}\, .
\label{KGSFT}
\end{equation} 
Exponential functions of the d’Alembertian operator naturally arise in string field theory, see e.g. \cite{koshelev,calcagni} and references therein. 

Motivated by the development of optomechanical experiments aimed at constraining non-local effects, we consider the paradigmatic case of the non-relativistic dynamics of a (1+1)-dimensional quantum harmonic oscillator. Following the approach described so far, we derive the nonlocal Schr\"{o}dinger equation
\begin{equation}
\label{nlse}
\mathcal{S} e^{\epsilon \mathcal{S}} \psi(t,x) = \frac{1}{2} m \omega^{2} x^2 \psi(t,x) \, ,
\end{equation}
where $m$ is the oscillator mass, $\omega$ its resonant frequency and $\epsilon = 2 m l_k^2/\hbar^2$. Notice that $\epsilon$ has dimensions of the inverse of an energy. By introducing the nonlocality energy scale as $E_k = \hbar c / l_k$, from the above definition of $\epsilon$ we see that $\epsilon E_k = 2 m c^2 / E_k$. The dimensionless quantity $\epsilon E_k$ gives (up to a factor 2) the ratio between the mass energy of the harmonic oscillator and the nonlocality energy scale. Thus $1/\epsilon$ is an energy scale directly related to the nonlocality energy scale when probed by an oscillator of mass $m$. For brevity, we will refer to $1/\epsilon$ simply as the \emph{nonlocality energy scale} from now on.

In order to cast Eq. (\ref{nlse}) in a dimensionless form amenable to theoretical and numerical analysis, we rescale the physical time as $\hat{t}=\omega t$, where $\omega$ is the angular frequency of the mechanical oscillator. 
The spatial coordinate $x$ is rescaled to $\hat{x}=x/x_0$, where $x_0 = \sqrt{\hbar/ m \omega}$. Using the dimensionless coordinates $\hat{t}=\omega t$ and $\hat{x}=x/x_0$, Eq.~(\ref{nlse}) takes the form 
\begin{equation}
\label{eq1}
\hat{\mathcal{S}} e^{\hat{\epsilon} \hat{\mathcal{S}}} \psi(\hat{t},\hat{x}) = \frac{1}{2} \hat{x}^2 \psi(\hat{t},\hat{x}),
\end{equation}
where $\hat{\mathcal{S}}=i \partial_{\hat{t}} + \frac{1}{2}\partial_{\hat{x} \hat{x}}^2$ and $\hat{\epsilon}=\epsilon \hbar \omega$. Eq.~(\ref{eq1}) reduces to the standard Schr\"odinger equation in the local limit $\hat{\epsilon}=0$. Note that $\sqrt{\hat{\epsilon}}$ corresponds to the ratio between $l_k$ and the amplitude of zero-point fluctuations $x_{ZPF} = x_0/\sqrt{2} = \sqrt{\hbar/ 2 m \omega}$ or, equivalently, $\hat{\epsilon}$ gives the dimensionless ratio between an energy quantum of the harmonic oscillator $\hbar \omega$ and the nonlocality energy scale $1/\epsilon$. Such dependence (as well as the above physical interpretation of $1/\epsilon$) suggests that massive quantum systems or, more precisely, systems with the smallest zero-point fluctuations could be the ideal setting for detecting these deviations. For a nonlocality length ranging between $l_k \sim 10^{-19}$m (higher values have been already excluded~\cite{biswas}) and the Planck scale $l_k \sim 10^{-35}$m, a quantum oscillator with $x_0 = 10^{-18}$m (see e.g. Ref.~\cite{bild}) would experience deviations from the ordinary Schr\"{o}dinger evolution ranging from $\hat{\epsilon} \sim 10^{-2}$ to $\hat{\epsilon} \sim 10^{-34}$. From now on, for simplicity, we shall omit hats in all quantities.

Assuming the usual factorization of the wavefunction $\psi(t,x)=\phi(x)exp(-i E t)$, where here $E$ is also rendered dimensionless dividing it by $\hbar \omega$, and using the fact that the exponential can be expanded as $\exp{(\epsilon \mathcal{S})}=(1 + \epsilon \mathcal{S} + \epsilon^2 \mathcal{S}^2/2! + \, \dots)$ we obtain the nonlocal time-independent Schr\"{o}dinger equation 
\begin{equation}
e^{\epsilon E} (E+\frac{1}{2} \partial_{xx}^2) e^{\frac{1}{2} \epsilon \partial_{xx}^2} \phi(x) = \frac{1}{2} x^2 \phi(x) \ .
\label{eq2}
\end{equation}
Eq.~\eqref{eq2} has a simpler interpretation in momentum space, where it takes the form
\begin{equation}
\label{eq2b}
\frac{1}{2} \frac{d^2 \tilde{\phi}(k)}{dk^2} + (E - V_{nl}(E,k))\tilde{\phi}(k) = 0 \ ,
\end{equation}
where $k$ is the dimensionless wavenumber ($k\rightarrow k \, x_0$), $\tilde{\phi}(k)$ is the Fourier transform of $\phi(x)$ and $V_{nl}(E,k)=E+(\frac{1}{2}k^2-E)e^{-\epsilon (\frac{k^2}{2}-E)}$. Hence, the nonlocal quantum evolution in a harmonic potential in real space is mapped into a standard, local quantum dynamics in an energy-dependent potential. The latter can be seen as encoding the residual backreaction of the quantum field on the background space. Interestingly, in the case of a quantum harmonic oscillator, the above correspondence holds for any analytic function $f$.

\section{Nonlocal quantum probabilities and the classical limit}

According to the correspondence principle, any model predicting deviations from ordinary quantum mechanics should consistently reproduce classical physics at high energies. In particular, for the case of the harmonic oscillator, the quantum mechanical probability density must tend to the classical probability density in the limit of large occupation numbers $n \rightarrow \infty$.
Before starting our analysis it is appropriate to clarify the meaning of ``classical limit'' of quantum dynamics (i.e. the high-energy/short-wavelength limit) in the presence of a nonlocality scale lying somewhere between the LHC TeV scale and the Planck scale. As discussed above, deviations from standard quantum mechanics are expected even at low energies, significantly below that defined by the nonlocality scale. A conceptually sound classical limit is thus defined for energies $E$ within the range $1/2 \ll E \ll 1/\epsilon$. This means energies significantly greater than the zero-point energy of the harmonic oscillator and yet notably lower than the nonlocality energy scale. It is in this regime that the classical limit associated with nonlocal modified quantum rules can be consistently compared with the standard one.

\subsection{Classical probability functions}
 
For a classical periodic system with total energy $E$ one can define the classical probability of measuring at an arbitrary time $t$ a wavenumber between $k$ and $k + dk$ as
\begin{eqnarray}
\rho(k)dk=\frac{1}{T}\frac{dk}{v(k)},
\label{ro}
\end{eqnarray}
where $v(k)$ is the oscillator velocity and $T$ is the half-period. By defining the turning points $k=\pm k_T$ as the wavenumber values at which the kinetic energy of the oscillator is zero, i.e. $k_T= \sqrt{2 E}$, we have
\begin{eqnarray}
T=\int_{-k_T}^{k_T}\frac{dk}{v(k)} \, .
\label{T}
\end{eqnarray}
The velocity $v(k)$ can be derived from the total energy relation $E = v^2/2 + V_{nl}$ giving 
\begin{eqnarray}
v(k)=\sqrt{(k_T^2-k^2)e^{-\frac{\epsilon}{2}(k^2-k_T^2)}}
\label{v}
\end{eqnarray}
Substituting Eq.~\eqref{v} in Eq.~\eqref{T} and integrating we obtain
\begin{eqnarray}
T=\pi \exp (-k_T^2\epsilon/8) \mathcal{I}_0(k_T^2\epsilon/8),
\label{Tsolved}
\end{eqnarray}
where $\mathcal{I}_j(z)$ is the $j$-order modified Bessel function of the first kind. For $\epsilon=0$, Eq.~\eqref{Tsolved} reduces to the local harmonic oscillator period $T = \pi$ ($\omega = 1$ in our dimensionless units). 
Using Eqs. (\ref{v}) and (\ref{Tsolved}) we can derive the classical probability density associated to the nonlocal harmonic oscillator as 
\begin{equation}
\rho(k)dk = \frac{e^{\frac{\epsilon}{4}(k^2-k_T^2/2)}}{\mathcal{I}_0(k_T^2\epsilon/8)} \rho_{loc}(k)dk,
\label{eq3}
\end{equation}
where $\rho_{loc}=1/(\pi \sqrt{k_T^2 - k^2})$ is the probability density function for $\epsilon=0$.
Similarly to the local case, the probability density (\ref{eq3}) goes to infinity at the turning points where the oscillator velocity is zero. In the classical limit, $k_T^2 \gg 1$, the probability of measuring wavenumbers $k \sim k_T$ is then substantially higher. 
For energies much smaller than the nonlocality energy scale, i.e. $k_T^2 \ll 1/\epsilon$, $\rho(k) \sim \rho_{loc}(k)$ thus reproducing the classical limit. 

The expectation values of the wavenumber $k$ in the local and nonlocal case are both zero due to the symmetry of the density functions about $k=0$. The expectation values of $k^2$ are instead finite and the following relation holds
\begin{equation}
\langle k^2 \rangle = \langle k^2 \rangle_{local} \left(1 + \frac{\mathcal{I}_1(k_T^2\epsilon/8)}{\mathcal{I}_0(k_T^2\epsilon/8)}\right).
\label{eq3b}
\end{equation}
For $k_T^2 \ll 1/\epsilon$, we find $\langle k^2 \rangle \sim \langle k^2 \rangle_{local}$. 
At energies comparable with the nonlocality scale, deviations reach up to 12 \%, indicating a gradual departure from classical predictions.

\subsection{Energy spectrum and quantum probability densities}

\begin{figure}
  \centering
\includegraphics*[width=1.\columnwidth]{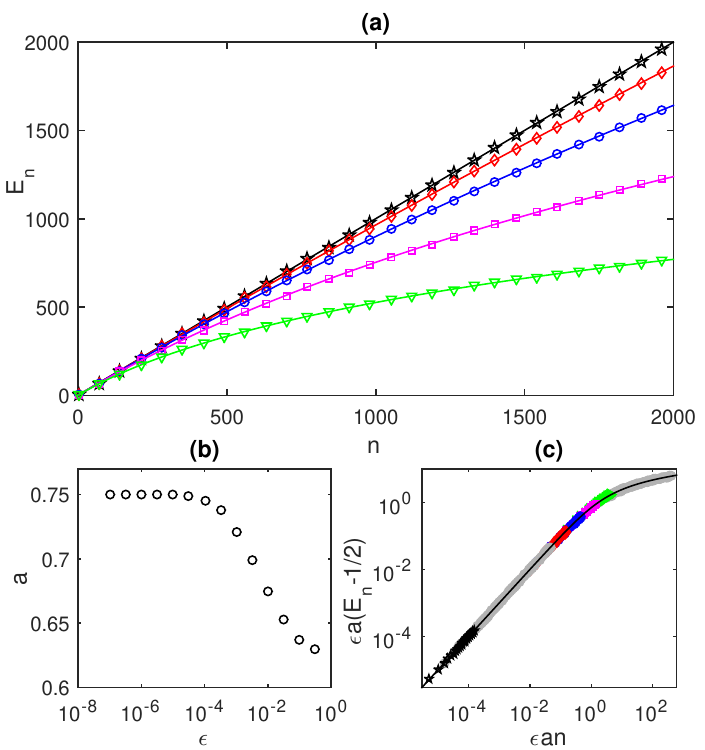}
\caption{a) Energy eigenvalues $E_n$ obtained from numerical integration of Eq.~\ref{eq2b} for $\epsilon=10^{-7}$ (black stars), $\epsilon= 10^{-4}$ (red diamonds), $\epsilon=3 \times 10^{-4}$ (blue circles), $\epsilon= 10^{-3}$ (magenta squares), $\epsilon=3 \times 10^{-3}$ (green triangles) and the corresponding fitting curves (solid lines) using Eq.~(\ref{eq4}) with $a$ as free parameter. b) Best-fit values of the parameter $a$ for different $\epsilon$. c) Collapse of the numerical data shown in a) onto the scaling function $E' \ln (1+n')$ (solid black curve), where $E'= \epsilon a (E_n -1/2)$ and $n'=\epsilon a n$. Additional data corresponding to other $\epsilon$ (see panel b)) are depicted in grey.}
\label{fig1}
\end{figure}

In order to calculate the quantum probability density functions both in the local and nonlocal cases and evaluate their convergence towards the classical predictions, we numerically integrate Eq.~(\ref{eq2b}) implementing a Numerov integration scheme and one-parameter shooting method based on bisection procedure to estimate the energy eigenvalues $E_n$. These are displayed in Fig.~\ref{fig1}a for different values of $\epsilon$.
Contrary to the local case, the eigenvalues show a nonlinear dependence on $n$ compatible with a logarithmic model. We apply a regression analysis using the fitting function 
\begin{equation}
E_n=\frac{1}{w(\epsilon) }\ln (1+ w(\epsilon) n)+ 1/2 \, ,
\label{eq4}
\end{equation} 
where $w(\epsilon)= a \epsilon$ leaving $a$ as free parameter. The values of $a$ obtained from the fits are reported in Fig.~\ref{fig1}b where error bars are omitted as they are smaller than the symbol size. These values show a weak dependence on $\epsilon$ reaching a plateau at approximately $a \sim 0.75$ for $\epsilon$ ranging from $10^{-4}$ up to $10^{-8}$. Eq.~(\ref{eq4}) suggests that the values of $E_n$ for different $\epsilon$ can be made to collapse on a single curve upon suitable rescaling of energies and quantum numbers as $E'= \epsilon a (E_n -1/2)$ and $n'=\epsilon a n$, for which Eq.~(\ref{eq4}) rewrites $E' = \ln (1+n')$ (see Fig.~\ref{fig1}c). The rescaled numerical data show an excellent agreement with the curve $E'(n')$ over a range of $n'$ of more than seven decades. For the $\epsilon$ values considered, the mean residuals vary from approximately $8 \times 10^{-13}$ for $\epsilon = 10^{-7}$ to $6 \times 10^{-2}$ for $\epsilon = 0.3$.

For $\epsilon=0$, Eq.~(\ref{eq4}) reduces to the local result $E_{loc}(n) = n + 1/2$. At the first order in $\epsilon$, i.e. for $n \ll 1/(a \epsilon)$, we find $E_n = E_{loc}(n) - a \epsilon n^2/2 + \mathcal{O}(\epsilon^2)$. This result can be compared to the one obtained in Ref.~\cite{belen19}, where first-order corrections to the energy eigenvalues of the nonlocal oscillator have been derived by means of time-independent perturbation theory giving $\mathcal{E}_n = E_{loc}(n) - \frac{3}{16}\epsilon (1+2n+2n^2)$. Using $a \approx 0.75$, we find a good agreement between the two results with deviations $(E_n-\mathcal{E}_n) \approx \frac{3}{8} \epsilon E_{loc}(n)$, i.e. smaller than $\epsilon$ at the ground state ($n=0$), and of the order of the zero-point energy at high occupation numbers $n \ll 1/(a \epsilon)$, where the first-order approximation of (\ref{eq4}) no longer holds.

\begin{figure}
  \centering
\includegraphics*[width=1.0\columnwidth]{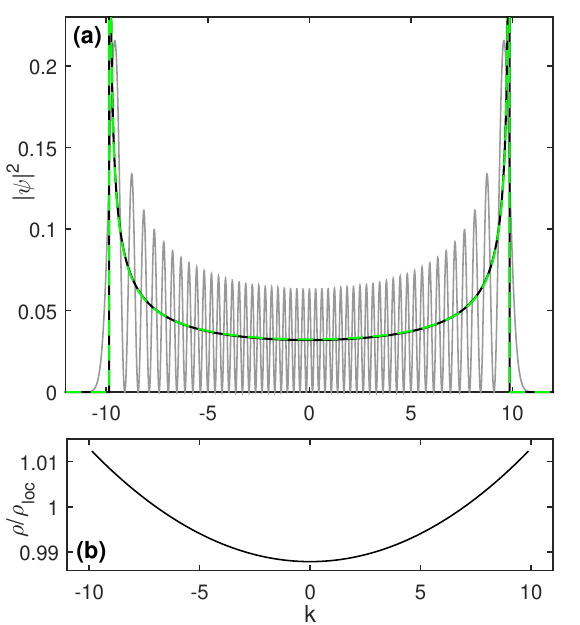}
\caption{a) Probability density function (solid grey) of the $n = 50$ state of the nonlocal quantum harmonic oscillator obtained by numerical integration of Eq.~(\ref{eq2b}) for $\epsilon = 10^{-3}$. The solid black (dashed green) line represents the local (nonlocal) classical probability function. b) Ratio of the classical probability densities $\rho(k)$ and $\rho_{loc}$ for the energy $E_{50}$.}
\label{fig2}
\end{figure}

An example of probability density for the $n = 50$ excited state of the nonlocal oscillator with $\epsilon=10^{-3}$ is shown in Fig.~\ref{fig2}a. The solid black and dashed green curves represent the classical probability densities $\rho_{loc}$ and $\rho$ evaluated for the energy $E_{50}$. For this energy, the two curves are similar within a margin of $1\%$, i.e. of the order of the zero-point fluctuations (see Fig.~\ref{fig2}b).

\begin{figure}
  \centering
\includegraphics*[width=1.0\columnwidth]{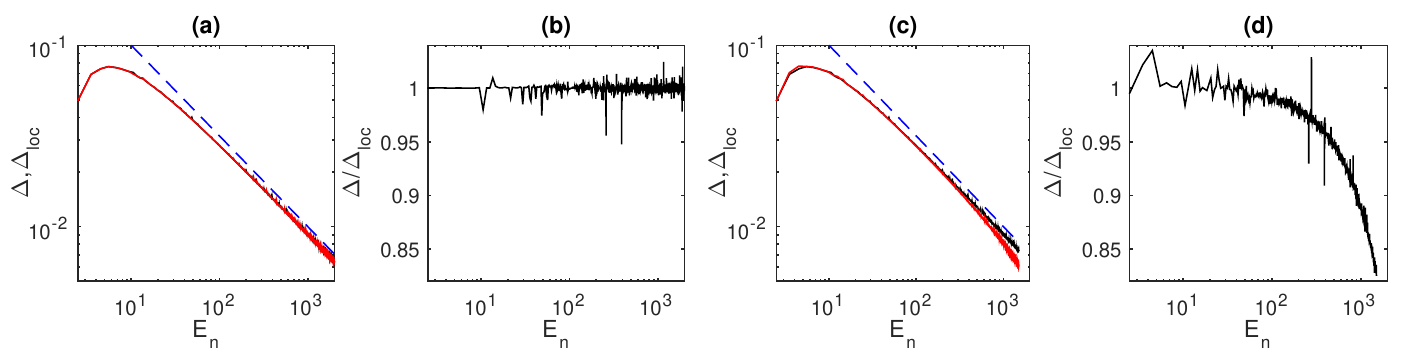}
\caption{Averaged difference between the maxima of the quantum probability and the classical probability densities $\Delta$ (red curve) and $\Delta_{loc}$ (black curve) defined in the text and their ratio for $\epsilon = 5 \times 10^{-7}$ (a-b) and $\epsilon = 5 \times 10^{-4}$ (c-d). The power law scaling $\sim E_n^{-1/2}$ is plotted for reference using dashed blue lines.}
\label{fig3}
\end{figure}

To characterize the convergence of the nonlocal quantum probability toward the classical limit, we define for each value of $n$ the quantity $\Delta$ as
\begin{equation}
\Delta (E_n) = \frac{1}{n+1} \sum _{i=1}^{n+1} (\vert \psi(k_i) \vert^2 - \rho_{loc}(k_i)) \, ,
\end{equation}
where the $k_i$ denotes the wavenumbers of relative maxima of $\vert \psi(k_i) \vert^2$ and $\rho_{loc}(k_i)$ is calculated using $k_T= \sqrt{2 E_n}$. This quantity represents the averaged difference between the maxima of the nonlocal quantum probability and the standard classical probability density $\rho_{loc}$. For comparison we also define $\Delta_{loc}$ calculated from (\ref{eq2b}) with $\epsilon=0$ and evaluated at similar energies $E_n$. 
We observe indeed that due to the nonlinearity of the energy spectrum (\ref{eq4}), similar energies $E_n$ in the local and nonlocal cases will correspond to different occupation numbers $n$.
In Fig.~\ref{fig3} we plot $\Delta$ and $\Delta_{loc}$ for two different values of $\epsilon$. For all energies considered in Figure \ref{fig3}a-b, it holds that $E_n \ll 1/\epsilon$: the maximum energy is indeed approximately $10^{3}$ times smaller of the nonlocality energy scale while it remains significantly larger than the ground state energy, thereby approaching a well-defined classical limit. The two curves in Fig.~\ref{fig3}a exhibit the expected power law scaling $\sim E_n^{-1/2}$ (dashed blue line), suggesting that both quantum probability densities converge toward the classical probability function, with $\Delta \approx \Delta_{loc}$ over the entire energy range (see Figure \ref{fig3}b).
On the other hand, deviations become apparent as energies approach the scale of nonlocality, as illustrated in Fig.~\ref{fig3}c-d, primarily due to the differing classical probability distributions $\rho$ and $\rho_{loc}$.

\section{Nonlocal Helmholtz free energy}

The simple structure of the energy levels allows us to compute some of the thermodynamic properties of the quantum harmonic oscillator, such as the partition function and the Helmholtz free energy. Using Eq.~\eqref{eq4} in the classical definition of the partition function we obtain
\begin{eqnarray*}
Z(\beta) = \sum _{n=0}^{\infty}e^{-\beta\left[(a\epsilon)^{-1}\ln(1+a\epsilon n)+\frac{1}{2}\right]}\\
=e^{-\frac{\beta}{2}}\sum _{n=0}^{\infty}e^{-\beta\left[(a\epsilon)^{-1}\ln(1+a\epsilon n)\right]}\\
=e^{-\frac{\beta}{2}}\sum _{n=0}^{\infty} (1+a\epsilon n)^{-\frac{\beta}{a\epsilon}}
\label{eq11}
\end{eqnarray*}
where $\beta=\hbar \omega / (k_B T_K)$, with $T_K$ the temperature and $k_B$ the Boltzmann constant.
Then, the free energy becomes

\begin{eqnarray}
F=-\frac{1}{\beta} \ln Z(\beta) &=&-\frac{1}{\beta}\left(\ln e^{-\frac{\beta}{2}}+\ln \sum _{n=0}^{\infty}(1+a\epsilon n)^{-\frac{\beta}{\epsilon a}}\right)\nonumber\\ 
&=& \frac{1}{2}-\frac{1}{\beta}\ln\sum _{n=0}^{\infty}(1+a\epsilon n)^{-\frac{\beta}{\epsilon a}}
\label{eq12}
\end{eqnarray}
The last sum in Eq.~(\ref{eq12}) converges for $\beta/(\epsilon a) > 1$. Since the parameter $a$ is $\mathcal{O}(1)$ (see Fig.~\ref{fig1}b), mathematical convergence occurs whenever the thermal energy of the oscillator is smaller than the nonlocality energy scale. This is consistent with the discussion in the previous section, where we argued that a meaningful classical limit could be defined only when the oscillator is not probing energy/length scales comparable with the nonlocality one.

We remind that for a local quantum harmonic oscillator, in the classical limit the difference between the quantum and the classical free energy is given by $F_{local}-F_{cl} = \frac{1}{2}$, where 
\begin{eqnarray}
F_{cl}=-\frac{1}{\beta} \ln Z_{cl}(\beta)=\frac{1}{\beta} \ln (\beta)
\label{eq9}
\end{eqnarray}

If this were to be the case also in the nonlocal scenario, the following relation would hold
\begin{eqnarray}
F-F_{cl}=\frac{1}{2}-\frac{1}{\beta}\ln \sum _{n=0}^{\infty}(1+a\epsilon n)^{-\frac{\beta}{\epsilon a}} -\frac{1}{\beta}\ln (\beta)=\frac{1}{2}
\label{eq10b}
\end{eqnarray}
implying
\begin{eqnarray}\label{Gseries1}
\sum _{n=0}^{\infty}(1+a\epsilon n)^{-\frac{\beta}{\epsilon a}}=\frac{1}{\beta}
\end{eqnarray}
The integral test for convergence applied to the series on the left hand side of Eq.~\eqref{Gseries1} 
gives the lower and upper bounds (see Appendix)
\begin{eqnarray}
\frac{1}{\beta-\epsilon a} \leq \sum _{n=0}^{\infty}(1+a\epsilon n)^{-\frac{\beta}{\epsilon a}} \leq 1+\frac{1}{\beta-\epsilon a}.
\label{limits1}
\end{eqnarray}
In the regime in which $1\ll 1/\beta \ll 1 /\epsilon$, i.e. for thermal energies much larger than the zero-point energy, but also much smaller that the nonlocality energy scale, Eq.~(\ref{limits1}) implies $\sum _{n=0}^{\infty}(1+a\epsilon n)^{-\frac{\beta}{\epsilon a}} =1/\beta$. Thus, we see that we retrieve the classical limit for the Helmholtz free energy of a local quantum harmonic oscillator.

\section{Quantum harmonic potential as a testbed for nonlocal theories}

{In the local case, the symmetry between position and momentum in the dynamics of the quantum harmonic oscillator implies that the Schr\"odinger equation has exactly the same form whether expressed in position or momentum space. For a non-local quantum oscillator, this symmetry no longer holds. Yet, when the nonlocal function is analytic, the relationship between the quadratic harmonic potential and the second derivative in momentum space allows the mapping of nonlocal quantum evolution in real space into a conventional local dynamics, but governed by an energy-dependent potential, in momentum space.}

{Indeed, the stationary Schr\"odinger equation for a generic (1+1)-dimensional nonlocal quantum oscillator
\begin{equation}
f(E+ \frac{1}{2} \partial_{xx}^2)\phi(x)= \frac{1}{2} x^2 \phi(x) \, 
\label{nlsgen}
\end{equation}
for any $f$ such that $f(z)=\sum_{j=1}^{\infty}b_j z^j$ 
with $b_1=1$, translates in momentum space into 
\begin{equation}
\frac{1}{2} \frac{d^2 \tilde{\phi}(k)}{dk^2} + (E - V_{nl}(E,k))\tilde{\phi}(k) = 0.
\label{nlsgen2}
\end{equation}
Eq.~\eqref{nlsgen2} is a stationary Schr\"odinger equation in an energy-dependent potential $V_{nl}(E,k)=E-f(E-k^2/2)$. The analyticity of the function $f$ is crucial here. In its absence nonlocality in real space would give rise to a nonlocal potential in momentum space involving convolution terms.

In the dimensionless units above defined, the nonlocal potential can be expanded as
\begin{equation}
V_{nl}(E,k)=\frac{k^2}{2} + \sum_{j=2}^{\infty}a_n \epsilon^{j-1}(E- \frac{k^2}{2})^j\\,
\label{potgen}
\end{equation}
namely a harmonic potential plus $\mathcal{O}(\epsilon)$-corrections. These corrections are responsible for the deviations from ordinary quantum mechanics and for those expected at high-energies $E \sim 1/\epsilon$. 
On the other hand, using similar arguments as those in Sec. III --- in particular that the energy dependent potential can be well approximated by the standard harmonic form $V_{nl}(E,k) \sim \frac{k^2}{2}$ for $k \sim k_T$ --- we could infer that the classical limit should be safely recovered. While these qualitative arguments cannot substitute a deeper analysis as the one conducted in this work, they suggest that our results could be generic of any analytic $f$, pointing towards harmonic oscillators as a valuable prototype model for investigating the classical limit of nonlocal field theories at non-relativistic energies.
\\

\section{Conclusions}

In this work we have considered the classical limit of a nonlocal quantum harmonic oscillator. The nonlocal model is derived from the non-relativistic limit of a d'Alambert operator with infinite many derivatives, inspired by string field theory. 
At the relativistic level, the nonlocal dynamics preserves Lorentz invariance, evading the rich observational constraints on Lorentz violations~\cite{liberati2013tests}. Furthermore, nonlocalities of different forms emerge in disparate quantum gravity models where they are central exactly in preserving Lorentz symmetry.

The non-relativistic limit allows us to assess the effect of nonlocality at energies characteristic of quantum mechanical systems amenable for investigation in the lab. Unlike previous works~\cite{belen16,belen17,belen19}, here we have tackled the problem at the non-perturbative level and focusing on the classical limit of the model.

We have shown that, for the specific case considered, the classical limit is non-pathological. We recover both the classical probability distribution for highly occupied states of the harmonic oscillator and a sensible thermodynamic behaviour looking at the Helmholtz free energy. At the same time, deviations appear, as expected, both in the low and high-energy limits. The former shows that quantum systems and experiments with massive objects are indeed a promising avenue for testing nonlocal effects. The latter, on the other hand, indicates that as we approach the energy scale associated with nonlocality, new physical effects are expected to arise.    

Although we focused on the non-relativistic limit of a specific non-local field operator, we believe that our results can be extended to other functional forms. For instance, this scenario likely holds for any analytic function $f$ where, as discussed before, the non-local quantum evolution in a harmonic potential in real space results in local quantum dynamics within an energy-dependent potential. 
On the other hand, even for operators for which the dynamics remains nonlocal also in Fourier space, the approach presented in this study --- based on the prototype model of a quantum harmonic oscillator --- may provide an useful tool for discerning pathological cases.


\section*{Acknowledgments}

AB acknowledges support from the Horizon Europe EIC Pathfinder project QuCoM (Grant Agreement No. 10104697).

\section*{Appendix}

Let us thus look at the convergence of the series 
\begin{eqnarray}
\sum _{n=0}^{\infty} G(n) = \sum _{n=0}^{\infty}(1+a\epsilon n)^{-\frac{\beta}{\epsilon a}}=\frac{1}{\beta}
\nonumber
\end{eqnarray}
The integral test for convergence applied to the infinite series gives the lower and upper bounds
\begin{eqnarray}
\int_{N}^{\infty}G(y)dy \leq \sum _{n=N}^{\infty}G(n)\leq G(N)+\int_{N}^{\infty}G(y)dy
\nonumber
\end{eqnarray}
With the change of variable $1+\epsilon ay=z$ we have $dy=\frac{dz}{\epsilon a}$. Then, setting $N=0$, the integral becomes 
\begin{equation}
\int_{0}^{\infty}G(y)dy=\int_{0}^{\infty}(1+\epsilon ay)^{-\frac{\beta}{\epsilon a}}dy=\frac{1}{\epsilon a}\int_{1}^{\infty}z^{-\frac{\beta}{\epsilon a}}dz
\nonumber
\end{equation}
from which we derive the lower bound
\begin{eqnarray}
\frac{1}{\epsilon a}\int_{1}^{\infty}z^{-\frac{\beta}{\epsilon a}}dz=\frac{1}{\beta-\epsilon a}
\nonumber
\end{eqnarray}
Since $G(y=0)=1$ we finally obtain
\begin{eqnarray}
\frac{1}{\beta-\epsilon a} \leq \sum _{n=0}^{\infty} G(n) \leq 1+\frac{1}{\beta-\epsilon a}.
\nonumber
\end{eqnarray}

\end{document}